# Temperature dependence of quasi-localized phonons-mediated non-Markovianity dynamics of $SiV^-$ centers in diamond

Wanggui Ye, Debao Zhang, Xuguang Cao, Ji Zhou, Xinye Fan, Sicheng Liu, Ke Yu, Jiqiang Ning, and Shijie Xu*

Department of Optical Science and Engineering, College of Future Information Technology, Fudan University, 2005 Songhu Road, Shanghai 200438, China

*Corresponding author, e-mail: xusj@fudan.edu.cn

**Abstract:** Here we investigate the temperature-dependent non-Markovian dynamics of the $SiV^-$ center in diamond, focusing on the roles of low- and high-frequency quasi-localized phonon modes. Low-frequency phonons exhibit stronger electron-phonon coupling, leading to long-lived dephasing rate, while high-frequency phonons induce rapid attenuation of oscillatory dephasing rate facilitating a persistent memory effect. The non-Markovianity measure $\mathcal{N}_C$ shows memory effects persisting at low temperatures but diminishing at high temperatures due to enhanced damping. The temperature dependence of $\mathcal{N}_C$ follows a monotonic decay, from which a transition temperature $T_{NM} = 110$ K is determined. These results highlight the interplay between phonon activation and damping in shaping quantum coherence, offering insights for optimizing solid-state quantum systems.

**Keywords:** $SiV^-$ Center, Diamond, Non-Markovianity, Electron-phonon coupling, Quasi-localized vibrational mode.

## I. INTRODUCTION

For applications such as quantum information processing, quantum computation, quantum cryptography, and high-precision quantum measurements, quantum systems must be shielded from deleterious environmental interactions through both passive isolation and active techniques [1], as interactions with the environment inevitably lead to decoherence. The negatively charged silicon-vacancy ($SiV^-$) color center in diamond is quickly emerging as an attractive candidate for these quantum technologies due to its high symmetry, deterministic creation, strong emission of the zero-phonon line (ZPL), and inherent robustness against environmental electric fluctuations [2,3]. Long coherence times, high-fidelity spin-photon interfaces, and scalability further enhance the potential of $SiV^-$ centers for quantum information storage and quantum sensing [4-6].

However, despite these significant advantages, the understanding of optical and quantum coherence properties of $SiV^-$ centers remain limited owing to the inevitable complicated interactions with phonons [7]. Electron-phonon coupling, particularly involving distinct quasi-localized vibrational modes, represents the major decoherence channel in these color centers. Such interactions yield rich and complex dynamics, extending well beyond traditional Markovian approximations commonly used in quantum optics [7]. In realistic solid-state quantum systems, the memory effects characterizing non-Markovian dynamics—manifested as a backflow of information from the environment to the quantum system—become pronounced [8]. These effects necessitate a deeper understanding and quantitative characterization, as standard Born-Markov approximations are insufficient [9], particularly in the presence of structured phononic environments, i.e., in the case of $SiV^-$ center.

Recently, significant theoretical progress has been made in quantitatively characterizing non-Markovianity (NM) from various perspectives, such as semigroup properties, divisibility criteria, coherence dynamics, and information backflow [10-13]. Nevertheless, experimental realizations and concrete demonstrations remain limited, especially regarding solid-state systems and their complex, structured environments. Moreover, the intricate relationship between electron-phonon coupling, phonon spectral densities, and non-Markovian memory effects in $SiV^-$ systems is still not fully understood. Bridging this theoretical-experimental gap is thus critical to advance the fundamental understanding of quantum decoherence in color centers.

In this article, we quantitatively investigate and report on the non-Markovian features inherent in the phonon-induced dephasing dynamics of $SiV^-$ color centers. Utilizing a coherence-based NM measure [14], we demonstrate explicit connections between phonon spectral densities, electron-phonon interactions, and non-Markovian behavior. Our results not only shed some light on the complex underlying physical mechanisms but also establish an experimental benchmark, bridging theoretical predictions and practical realizations.

## II. THEORETICAL METHODS

### A. Quantifying Non-Markovianity

Following microscopic Hamiltonian [5,15] is considered to describe the local interaction of a qubit and a bosonic thermal reservoir for $SiV^-$ centers in diamond (in units of $\hbar$),

$$H = \omega_0 \sigma_z + \sum_k \omega_k a_k^\dagger a_k + \sum_k \sigma_z \big(g_k a_k + g_k^* a_k^\dagger\big), \tag{1}$$

where $\omega_0$ is the bare transition frequency of the qubit system, $\omega_k$ is the frequency of $k$th quasi-localized phonon mode, $a_k$ ($a_k^\dagger$) is the annihilation (creation) operators, and $g_k$ is the coupling constant between each phonon mode and the qubit system, which arises from the different deformation potentials of the ground and the excited state. In the continuum limit, $\sum_k |g_k|^2$ could be transformed to $\int d\omega\, \mathcal{J}(\omega)\delta(\omega_k - \omega)$, where $\mathcal{J}(\omega)$ is the reservoir's spectral density function (SDF) [16]. Thus, the exact dynamics of the $SiV^-$ center resulting from Eq. (1) is described by the master equation in the interaction picture [14,17],

$$\frac{d\rho_s}{dt} = -\frac{\gamma(t)}{2}[\rho_s(t) - \sigma_z\rho_s(t)\sigma_z]. \tag{2}$$

It is assumed that the phonon reservoir is initially in a thermal state [9] $\rho_{ph} = \exp(-\beta H_{ph})/Tr\{e^{-\beta H_{ph}}\}$, where $\beta = (k_B T)^{-1}$ and $H_{ph} = \sum_k \omega_k a_k^\dagger a_k$. Thus, the time-dependent dephasing rate which fully determines the system-environment interaction is given by [15],

$$\gamma(t) = \int_0^\infty d\omega \frac{\mathcal{J}(\omega)}{\omega} \coth\left(\frac{\hbar\omega}{2k_B T}\right)\sin(\omega t), \tag{3}$$

giving rise to the evolution of the density matrix off-diagonal elements $\rho_{ij}(t) = e^{-\Gamma(t)}\rho_{ij}(0)$, $i \neq j$, with dephasing factor $\Gamma(t) = 2\int_0^t \gamma(\tau)d\tau$. And for this model the coherence has the form [14],

$$C(t) = 2|\rho_{ij}(0)|e^{-\Gamma(t)} \tag{4}$$

As coherence is monotonically reducing for $t \geq 0$ under incoherent completely positive and trace preserving maps for Markovian dynamics, any violation of this monotonicity at any time, i.e., $dC/dt > 0$, will be served as an NM indication. Therefore, the following measure [14] is used to quantifying the NM for such incoherent open system dynamics,

$$\mathcal{N}_C = \max_{\rho(0)\in\mathcal{F}^C} \int_{dC(\rho(t))/dt>0} \frac{dC(\rho(t))}{dt} dt$$
$$\xrightarrow{\rho_{ij}(0)=1/2} -2\int_{\gamma(t)<0} \gamma(t)e^{-\Gamma(t)}dt. \tag{5}$$

Note that the integral constraints are equivalent for this case. They indicate that the coherence NM measure is intrinsically related with the backflow of information in system-reservoir interaction, which results in a temporal increment of coherence of the system. Eq. (5) establishes an association between the NM onset and the phononic spectral density function, such that the effect of quasi-localized phonon models on $SiV^-$ optical dynamics in a realistic diamond lattice may be elucidated.

### B. Photoluminescence Spectrum and Spectral Density Function

In terms of the critical role of the spectral density function $\mathcal{J}(\omega)$ of phonons in the electron-phonon coupling and hence coherence NM measure, its profile will be determined from the line shapes of photoluminescence (PL) spectrum of $SiV^-$ center in diamond. Like the case of $NV^-$ center [18], the PL spectrum of $SiV^-$ center comprises the ZPL lines and accompanying phonon sidebands due to the interactions between quasi-localized vibrations and electronic states.

An optical transition between two energy levels under the action of an external optical field can be elaborated by a generating linear optical polarization, i.e., $P^{(1)}(t) = \varepsilon_0 \chi^{(1)} E(t)$, where $\varepsilon_0$ is vacuum permittivity, $\chi$ susceptibility, and $E(t)$ the electric field component of the incident electromagnetic field [19]. Usually, in the analysis of time-domain spectroscopic measurements, the polarization is reformulated utilizing optical response functions $S^{(1)}$ to facilitate theoretical interpretation and computational modeling, i.e., $P^{(1)}(t) = \int_{-\infty}^{t} S^{(1)}(t_1) E(t - t_1) \, dt_1$. Note that within the framework of time-domain spectroscopy, the linear optical response function can be physically interpreted as characterizing the temporal evolution of the system's density matrix, which undergoes sequential transformations through interactions with each incident optical field [20]. Thus, the linear optical response function could be written in the form: $S^{(1)}(t) = \frac{i}{\hbar} e^{-i\omega_{eg} t - g(t)} + c.c.$, which reflects the dynamics of the coherence of the two energy states involving the dephasing line shape function $g(t)$. And the frequency-domain fluorescence spectrum [21] is given by Fourier transform of the optical response:

$$\sigma_f(\omega) = \frac{1}{\pi} Re \int_0^\infty e^{i(\omega - \omega_{eg} + \lambda)t - g^*(t)} dt, \tag{6}$$

where $\lambda$ is the vibrational reorganization energy resulting from the effect of lattice vibration on electronic states. For the spin-boson model mentioned above, the dephasing line shape function can be written as [22]:

$$g(t) = -\frac{1}{2\pi} \int_{-\infty}^{+\infty} d\omega \mathcal{J}(\omega) \left[1 + \coth\left(\frac{\beta \hbar \omega}{2}\right)\right] \left[e^{-i\omega t} + i\omega t - 1\right], \tag{7}$$

$$Re[g(t)] = \frac{1}{2\pi} \int_{-\infty}^{+\infty} d\omega \mathcal{J}(\omega) \coth\left(\frac{\beta \hbar \omega}{2}\right) [1 - \cos(\omega t)], \tag{8}$$

$$Im[g(t)] = \frac{1}{2\pi} \int_{-\infty}^{+\infty} d\omega \mathcal{J}(\omega) [\sin(\omega t) - \omega t], \tag{9}$$

where $\beta = (k_B T)^{-1}$.

Previous studies have identified two prominent quasi-localized vibrational modes (QLMs) with characteristic energies around 45 meV and 63 meV that are strongly involved in electron-phonon interactions for color centers in diamond, notably $NV^0$ and $NV^-$ centers [18]. In the $SiV^-$ center system discussed here, analogous vibrational modes, including the breathing mode (symmetry $a_{1g}$, labeled QLM1) and the silicon-atom-related mode ($a_{2u}$, labeled QLM2), play a crucial role in determining the spectral density function (SDF). Accordingly, the total SDF for $SiV^-$ centers can be expressed as a sum of contributions from these two QLMs: $\mathcal{J}(\omega) = \sum_{i=1,2} \mathcal{J}_i(\omega)$, with each component typically characterized by a Lorentzian-like spectral densities distribution [23, 24]:

$$\mathcal{J}_i(\omega) = \frac{2\lambda_i \omega_i^2 \omega \xi_i}{(\omega \xi_i)^2 + (\omega_i^2 - \omega^2)^2}, \tag{10}$$

where $\omega_i$ is the specific localized phonon frequency, $\xi_i$ is the characteristic width, and $\lambda_i$ is the $ith$ mode's contribution to the vibrational reorganization energy, which is positively related to Huang-Rhys factor $S$ intrinsically reflecting the coupling strength of electron-phonon.

## III. RESULTS AND DISCUSSION

A high-quality diamond bulk sample was cooled to 10 K in a helium gas closed-cycle cryostat, and optically measured on a home-assembled multi-function integrated micro-spectroscopy setup equipped with a 532 nm CW laser as the excitation light resource. As illustrated in Fig. 1(a), the $SiV^-$ center in diamond consists of a silicon atom occupying an interstitial site between two adjacent carbon vacancies aligned along the [111] crystallographic axis. This configuration forms a split-vacancy structure, where the silicon atom is symmetrically bonded to two triangular carbon sublattices on either side. As a result, the $SiV^-$ center possesses inversion symmetry ($D_{3d}$ point group). Such symmetric lattice configuration plays a crucial role in the optical selection rules, electronic structure, and phonon coupling behavior of $SiV^-$ center. The electronic level structure of the $SiV^-$ center consists of doubly degenerate ground and excited state orbitals, as shown in Fig. 1(b). However, inclusion of spin-orbit coupling lifts the degeneracy, leading to energy splitting of approximately ~0.3 and 1.1 meV [25], respectively. Therefore, at cryogenic temperatures, the ZPL line of $SiV^-$ center generally consists of four sharp lines labeled as A, B, C and D. Besides, trace of $Si^{29}$ and $Si^{30}$ isotopes in the studied sample also produces weak structures in the PL spectrum: the former for $A' - D'$, while the latter for $A'' - D''$, as depicted in Fig. 1(d). The fine structure splitting remains almost identical across different isotopes, whereas the whole spectrum shifts toward the longer wavelength direction as previously observed [26].

Figure 1(c) shows measured 10 K PL spectrum of the $SiV^-$ center in the studied diamond, covering the wavelength range of 730-790 nm. The two phonon structures at 45.5 and 63.9 meV, relative to the ZPL line (C peak), can be well resolved. As mentioned earlier, these modes correspond to the $a_{1g}$ and $a_{2u}$ symmetry representations, respectively. Their atomic displacement patterns are illustrated in Fig. 1(a), which depicts two representative QLMs associated with the negatively charged $SiV^-$ defect. In QLM1 (left panel), the vibrational energy is primarily distributed over the nearest-neighbor carbon atoms which undergo transverse, in-plane displacements relative to the $SiV^-$ axis. The silicon atom remains nearly stationary, indicating a breathing-like motion of the surrounding lattice. In contrast, QLM2 (right panel) exhibits a pronounced out-of-plane component: the silicon atom oscillates along the [111] axis between the two adjacent carbon vacancies, while the surrounding carbon atoms vibrate in a coordinated axial fashion. This shear-like mode reflects the inversion symmetry of the $SiV^-$ center and is more strongly localized at the defect core. Both modes demonstrate partial localization of vibrational energy near the defect, with distinct spatial symmetries and atomic contributions.

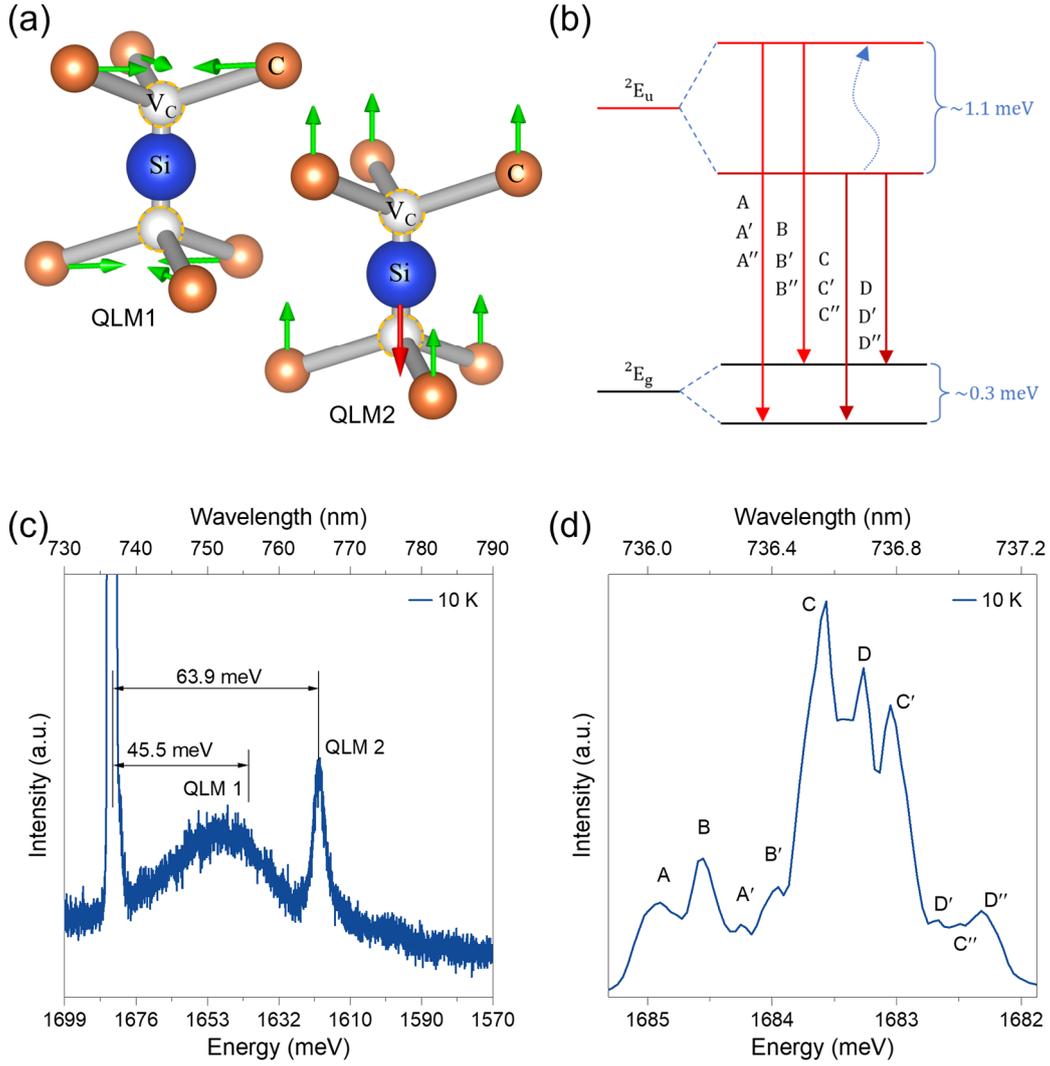

Figure 1 (a) Schematic atomic structure of SiV⁻ color center in diamond. The green and red arrows indicate the vibrational directions of the carbon and silicon atoms, respectively, illustrating two distinct quasi-localized vibrational modes: QLM1 (left) and QLM2 (right). (b) Energy level diagram of SiV⁻ center. Four arrows denote the optical transitions labeled A (A′, A″), B (B′, B″), C (C′, C″) and D (D′, D″) of $Si^{28}$ ($Si^{29}$, $Si^{30}$). (c) Magnified 10 K PL spectrum showing two QLM structures. (d) 10 K ZPL spectrum of Si-isotopes in the studied diamond sample.

    Based on the theory model outlined in Section II(B), the variable-temperature PL spectra of the SiV⁻ center have been simulated. In the theoretical spectra, the main spectral structures including the ZPL as well as the phonon sidebands are well reproduced. Note that the two QLMs are considered to couple with all the optical transitions including ones of isotopes, and the transition dipole moments $\mu_{ij}$ are incorporated into the model during the simulation process to account for the relative intensity weights between the ZPLs. Normalized PL spectra (dotted curves) measured at 10 K, 110 K, and 210 K are shown in Figures 2(a), (b), and (c) respectively. Various calculated ZPL lines (colorful filled peaks) and their cumulative line (solid line) are

also shown in the figures. Good agreement between experiment and calculation is achieved. As the temperature increases, the individual ZPLs gradually broaden and become indistinguishable. As shown in the upper panel figures, the two QLM structures of the SiV⁻ center can be well resolved at 10 K, but become progressively indistinguishable at higher temperatures. The theoretical spectra also show such tendency with temperature.

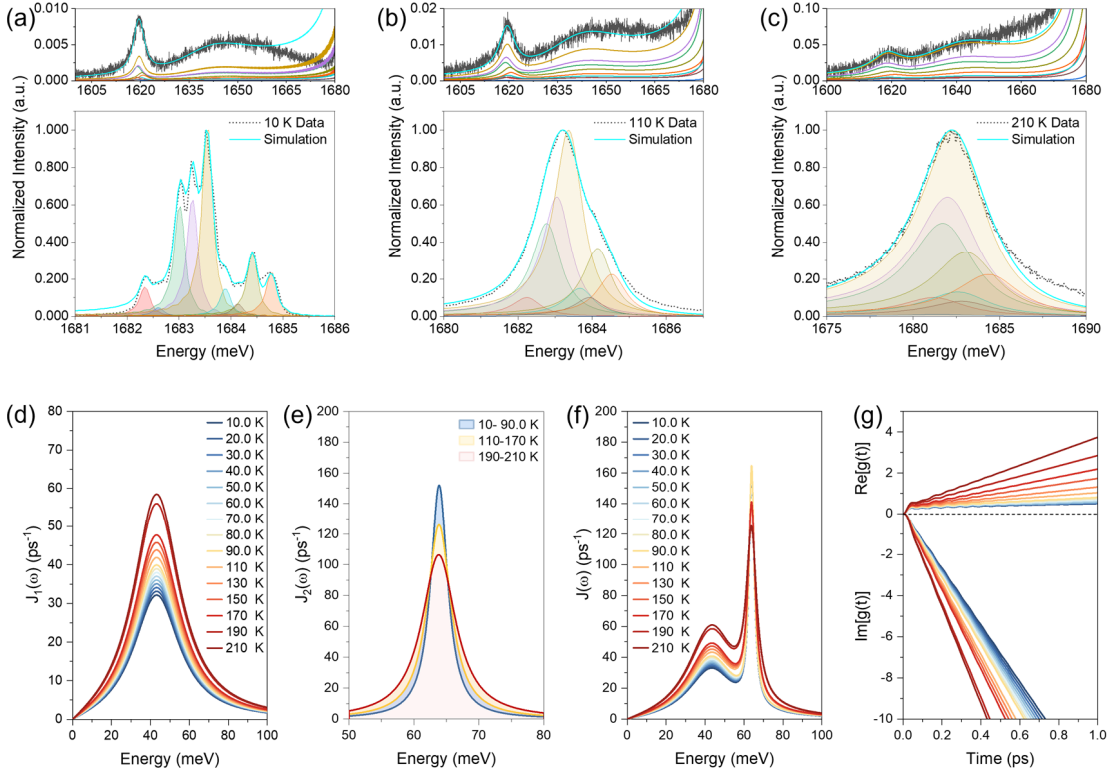

Figure 2. Calculated (various colorful filled peaks and cumulative solid curve) and experimental (dotted curve) PL spectra at (a) 10 K, (b) 110 K and (c) 210 K. Two QLM structures can be well resolved in the PL spectra shown in upper panel, especially at low temperatures. Obtained individual SDF profiles of (d) the QLM1 (low-frequency), (e) QLM2 (high-frequency), and (f) total ones at different temperatures. (g) Real part (upper panel) and imaginary part (bottom panel) of the line shape function $g(t)$ for various temperatures. Note that the legend of (g) is consistent with (f).

The SDF profiles $\mathcal{J}_1(\omega)$ and $\mathcal{J}_2(\omega)$ for QLM1 and QLM2 at various temperatures, determined from the simulations to the experimental PL spectra, are shown in Figs. 2(d) and 2(e), respectively, with the total SDF $\mathcal{J}(\omega)$ presented in Fig. 2(f). As temperature increases, $\mathcal{J}_1(\omega)$ continuously rises, whereas $\mathcal{J}_2(\omega)$ decreases in three distinct stages within the interested temperature range. This contrasting behavior indicates that electron-phonon coupling of SiV⁻ center involving the low-frequency QLM1 is more sensitive to temperature compared to the high-frequency QLM2. This can be explained by analyzing temperature-dependent parameters in Table 1: for QLM1, the coupling strength $\lambda_1$ steadily increases with constant damping $\xi_1$, thereby facilitating phonon propagation and electron-phonon coupling. Conversely, in QLM2, the damping parameter $\xi_2$ markedly increases with temperature, surpassing the relatively slower increase of coupling strength $\lambda_2$. Hence, the denominator of Eq. (10) grows faster than the numerator, causing $\mathcal{J}_2(\omega)$ to decrease. This increased damping accelerates energy dissipation, promotes phonon localization, and thus reduces effective coupling. Additionally, low-energy phonons (QLM1) are more readily thermally excited, substantially intensifying their interaction with electronic states, whereas high-energy phonons (QLM2), being less thermally populated and increasingly localized, contribute less effectively. Consequently, as temperature rises, low-frequency phonons dominate the vibrational renormalization energy, as reflected by the rapid increase of $\lambda_1$ shown in Table 1.

Table 1. Key simulation parameters for the two QLM in the PL spectra at various temperatures.

| | 10 K | 30 K | 50 K | 70 K | 90 K | 110 K | 130 K | 150 K | 170 K | 190 K | 210 K |
|---|---|---|---|---|---|---|---|---|---|---|---|
| $\lambda_1$ | 6.6 | 7.0 | 7.4 | 7.8 | 8.2 | 8.6 | 9.0 | 9.4 | 9.8 | 11.5 | 12 |
| $\xi_1$ | $0.64\omega_1$ | $0.64\omega_1$ | $0.64\omega_1$ | $0.64\omega_1$ | $0.64\omega_1$ | $0.64\omega_1$ | $0.64\omega_1$ | $0.64\omega_1$ | $0.64\omega_1$ | $0.64\omega_1$ | $0.64\omega_1$ |
| $\lambda_2$ | 2.5 | 2.5 | 2.5 | 2.5 | 2.5 | 2.9 | 2.9 | 2.9 | 2.9 | 3.5 | 3.5 |
| $\xi_2$ | $0.05\omega_2$ | $0.05\omega_2$ | $0.05\omega_2$ | $0.05\omega_2$ | $0.05\omega_2$ | $0.07\omega_2$ | $0.07\omega_2$ | $0.07\omega_2$ | $0.07\omega_2$ | $0.1\omega_2$ | $0.1\omega_2$ |

Figure 2(f) shows real and imaginary parts of $g(t)$ as a function of time at different temperatures. The contrasting temperature dependences of $\mathcal{J}_1(\omega)$ and $\mathcal{J}_2(\omega)$ result in asymmetric growth rates of the real and imaginary components of $g(t)$. The line shape function $g(t)$, which captures the dynamics of electron-phonon interactions, provides insights into the distinct dynamics of low-frequency and high-frequency phonons, such as revealing how these modes contribute differently to electron-phonon interactions. The real part of $g(t)$, characterized by the $\mathcal{J}(\omega)\coth(\beta\hbar\omega/2)$ term, which is rooted in the fluctuation-dissipation theorem, reflects the influence of thermal fluctuations. As temperature increases, this term amplifies the contribution of low-frequency phonons (QLM1), whose coupling strength $\lambda_1$ grows continuously while the damping parameter $\xi_1$ remains constant, and consequently, leads to a steady increase in the real part of $g(t)$. This trend suggests that low-frequency phonons play a critical role in modulating the phase of electronic states, contributing to the vibrational renormalization energy and maintaining coherent electron-phonon coupling. In contrast, the imaginary part of $g(t)$, subjected to the linear term $-\omega t$, reflects rapid energy dissipation due to the scattering from high-frequency phonons (QLM2). As temperature increases, the damping parameter $\xi_2$ rises significantly, leading to stronger localization effects and faster energy loss. This causes the imaginary part to decrease steeply, suggesting that high-frequency phonons contribute slightly to long-term electron-phonon coupling due to their rapid dissipation.

The steeper slope of the imaginary part compared to the real part highlights a distinct separation in the roles of low-frequency and high-frequency phonons: the former enhance coupling through thermal fluctuations, while the latter act primarily as channels for energy dissipation.

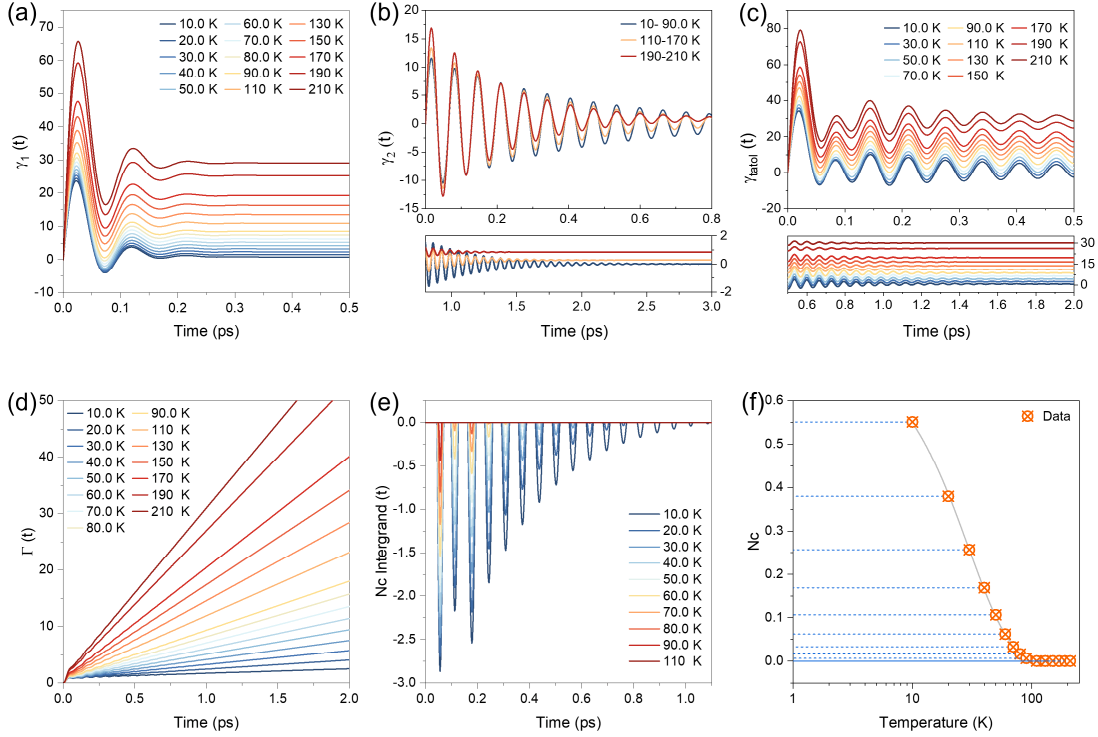

Figure 3. Dephasing rate $\gamma$ for various temperatures by considering the contribution of each SDF: (a) $\mathcal{J}_1(\omega)$ of low-frequency QLM1, (b) $\mathcal{J}_2(\omega)$ of high-frequency QLM2, and total SDF (c) $\mathcal{J}(\omega) = \sum_{i=1,2} \mathcal{J}_i(\omega)$. (d) Dephasing factor calculated based on the total dephasing rate for different temperatures. (e) Curves of the integrand of $\mathcal{N}_C$ integral function for various temperatures. (f) Plot of $\mathcal{N}_C$ values for quantifying the NM versus temperature.

The system-environment interaction and thus the NM degree is fully determined by the time-dependent dephasing rate $\gamma(t)$ described by Eq. (3). Figure 3(a) shows temperature-dependent behavior of $\gamma_1(t)$, driven by the low-frequency QLM1 phonons. It exhibits a pronounced oscillatory behavior with a rapid initial increase followed by quickly damped oscillations across the interested temperature range of 10-210 K. The initial oscillatory amplitude of $\gamma_1(t)$ increases significantly with the increasing temperature. This is a direct consequence of the enhanced phonon occupation numbers at elevated temperatures, as described by the $\coth(\beta\hbar\omega/2)$ term. On the other hand, the temperature-dependent behavior of $\gamma_2(t)$, which reflects the dephasing contributions from the high-frequency QLM2 phonons, exhibits distinct oscillatory patterns across different temperature ranges, as shown in Fig. 3(b). The curves are categorized into three segments based on temperature: 10-90 K (blue), 110-170 K (orange), and 190-210 K (red). Unlike $\gamma_1(t)$, the oscillatory amplitude and frequency of $\gamma_2(t)$ show relatively weaker dependence on temperature. This also suggests that the high-frequency phonons possess relatively weaker thermal population effects compared to the low-frequency phonons.

As time progresses, both $\gamma_1(t)$ and $\gamma_2(t)$ converge towards their respective steady states. Note that in line with the SDF, the total dephasing rate of the $SiV^-$ center defined in Eq. (3) is written as $\gamma(t) = \sum_{i=1,2} \gamma_i(t)$. Therefore, a noticeable feature of $\gamma_{total}(t)$ is the upward shift of the long-time oscillatory baseline as temperature increases, particularly significant for $\gamma_1(t)$, as shown in Fig. 3(c). In contrast, $\gamma_2(t)$ associated with the high-frequency phonons shows minimal baseline shift and remains close to zero in the long-time limit, as seen in the bottom panel figure in Fig. 3(b). Consequently, the steady-state dephasing rate is predominantly governed by the thermally activated low-frequency phonons, which maintain a higher population and a larger coupling strength $\lambda_1$ at steady state, as evidenced in Fig. 3(c). This distinct behavior implies that the low-frequency phonons are the primary drivers of long-term dephasing and coherence loss, while the high-frequency phonons primarily contribute to the short-term oscillatory dephasing of the $SiV^-$ center. Additionally, integrating $\gamma_{total}(t)$ yields the cumulative dephasing factor $\Gamma(t)$, depicted in Fig. 3(d). Clearly, $\Gamma(t)$ exhibits a nearly linear increase at long times, with a significantly steeper slope at higher temperatures, further confirming that enhanced phonon activation at elevated temperatures intensifies overall coherence loss.

The complementation between low- and high-frequency phonon modes underscores the complexity of phonon-mediated decoherence in the $SiV^-$ center, particularly through NM dynamics characterized by negative regions in $\gamma(t)$. At lower temperatures, deeper and extended negative dips observed in $\gamma_1(t)$ emphasize pronounced NM effects due to reduced phonon populations, prolonged coherence, and less rapid information dissipation. Conversely, the consistent magnitude and duration of negative dips in $\gamma_2(t)$ across all temperatures indicate persistent memory effects arising primarily from coherent oscillatory exchanges driven by QLM2 rather than thermal processes. Specifically, the early-time regime (before 0.2 ps) for $\gamma_2(t)$ shows increasing oscillation amplitudes with temperature, suggesting enhanced initial phonon-electron interactions facilitated by thermally activated phonons. Beyond this timescale, the trend reverses, as increased damping ($\xi_2$) at higher temperatures rapidly dissipates coherence, weakening oscillations. This amplitude reversal reveals a competition between initial phonon population-induced coherence loss and subsequent damping-driven dissipation. Thus, the transient information backflow enabled by phonon interactions is swiftly suppressed at higher temperatures. These observations illustrate the significant differences in the contributions of phonon modes QLM1 and QLM2 to NM dynamics in $SiV^-$ centers.

Figure 3(e) presents the integrand of the NM measure $\mathcal{N}_C$ as a function of time at various temperatures, calculated from Eq. (5), which quantifies the time-resolved contribution to the cumulative NM value. A clear trend can be found: as temperature increases, the integrand amplitudes diminish more rapidly with time, and their oscillatory contributions become confined to the early-time region. This behavior reflects the interplay between phonon populating and damping effects. At elevated temperatures, enhanced damping shortens the memory effect duration and suppresses long-time contributions to $\mathcal{N}_C$. Conversely, at lower temperatures (blue curves), reduced damping permits longer-lasting oscillations, sustaining the contributions to $\mathcal{N}_C$ over extended timescales. Consequently, the integrated NM measure value $\mathcal{N}_C$, shown in Fig. 3(f), exhibits a monotonic decrease with increasing temperature. A critical temperature $T_{NM} = 110$ K is identified, marking the transition from a non-Markovian to a predominantly Markovian regime. This result highlights the dominant role of phonon-induced damping in suppressing memory effects at elevated temperatures, ultimately leading to a progressive reduction in information backflow within the system-environment interaction.

## IV. CONCLUSIONS

In conclusion, this study reveals the distinct roles of low- and high-frequency quasi-localized phonon modes in electron-phonon coupling and non-Markovian dynamics of the $SiV^-$ center. Low-frequency QLM1 phonons exhibit stronger electron-phonon coupling, leading to significant long-term dephasing that intensifies with temperature due to enhanced phonon occupation. In contrast, high-frequency phonons contribute primarily to short-lived oscillatory dephasing, with their influence being less temperature-sensitive. The dephasing rates $\gamma_1(t)$ and $\gamma_2(t)$ highlight a competition between phonon populating and damping effect, with an observed amplitude reversal over time. Therefore, the NM measure $\mathcal{N}_C$ confirms that a persistent memory effect is facilitated by high-frequency phonons. The memory effects are more prominent at lower temperatures, where weaker damping sustains coherence for longer durations. As temperature rises, damping dominates, confining NM contributions to short timescales. And the temperature dependence of $\mathcal{N}_C$ follows a monotonic decay, defining a characteristic transition temperature $T_{NM} = 110$ K, beyond which non-Markovian effects become negligible. These findings accentuate the interplay between phonon mode, damping effect, and electron-phonon coupling strength in determining the $SiV^-$ center's non-Markovian properties. The results may provide critical insights into optimizing quantum NM in solid-state defect systems by controlling phonon interactions via phonon engineering.

**Acknowledgments:** This work was financially supported by the National Science Foundation of China (No. 12074324).